\documentclass[11pt,twosided]{article}
\usepackage{newpasp,psfig}
\reversemarginpar

\def\pc{\,{\rm pc}}\def\kpc{\,{\rm kpc}}\def\Mpc{\,{\rm Mpc}}
\def\kms{\,{\rm km}\,{\rm s^{-1}}}\def\msun{\,{\rm M}_\odot}
\def\W{\,{\rm W}}\def\y{\,{\rm y}}\def\kev{\,{\rm keV}}
\def\fracj#1#2{{\textstyle{#1\over#2}}}

\index{Binney, J. J.}

\begin{document}
\title{Supernovae and the IGM}
 \author{James Binney}
\affil{Oxford University, Theoretical Physics, Keble Road, Oxford, OX1 3NP,
U.K.}

\begin{abstract}
An energetic argument implies that a galaxy like the Milky Way is blowing a
powerful wind that carries away most of the heavy elements currently
synthesized and has impacted the IGM out to at least $180\kpc$.  Rich
clusters of galaxies appear to be closed systems in which
most heavy elements are ejected from galaxies. More supernovae are required
than the yield of core-collapse SNe from a Salpeter IMF. X-ray observations
imply that the IGM in groups and clusters as been strongly preheated. SNe probably cannot supply
the required energy, which must come from AGN. 
\end{abstract}

\section{Introduction}

Observations of both cool and hot intergalactic gas make it clear that heavy
elements are by no means confined to galaxies. 

Larson (1974) predicted that late in the galaxy formation process, galactic
winds would carry heavy elements out into intergalactic space. Soon
afterwards X-ray spectra of intracluster gas confirmed that the gas was quite
metal-rich.  In recent years observations of absorption lines in QSO spectra
have elucidated the impact of supernovae on the diffuser gas that is found
in the field, at least at high redshift. These observations imply that, from
a redshift $\ga3$ onwards, there does not seem to be a parcel of gas that
has not felt the impact of spernovae, probably mediated by galactic winds.

While the ubiquity of the products of supernovae is beyond doubt, many
aspects of the interaction of supernovae and intergalactic gas are highly
uncertain. Outstanding questions include: (i) in what types of galaxy were
intergalactic metals manufactured? (ii) which type of supernova
(core-collapse or Ia) has been most important, and at what epoch? (iii) how
important have supernovae been for the entropy budget of intergalactic gas?

\section{Winds from spiral galaxies}

Most of the luminosity in the Universe comes from spiral galaxies like the Milky
Way. Do such galaxies blow SN-driven winds?

Studies of the local ISM show that a significant fraction of the pressure in
the ISM comes from cosmic-rays and the magnetic field. Supernovae and fast
stellar winds are the main energy sources of the ISM, so the dynamical
importance of cosmic rays and magnetic fields suggests that a significant
fraction of the kinetic energy which these objects pump into the ISM is
chanelled into cosmic rays and magnetic fields. Hence, assuming there to
be $\sim3$ supernova per century, the cosmic-ray reservoir is being
energized at a rate $10^{44}/3\times10^9\simeq3\times10^{34}\W$. How does
this power, which could build up the observed interstellar pressure within a
disk $8\kpc$ in radius and $200\pc$ thick within $\sim10^5\y$, manifest
itself?  

The local cosmic-ray energy spectrum is such that the energy is
overwhelmingly contained in only mildly relativistic particles.
Such particles have enormous lifetimes because they are too fast to suffer
much Coulomb scattering and too slow to produce significant synchrotron
radiation -- which explains why the non-thermal radio luminosity of galaxies
like the Milky Way is $\sim10^{30}\W$ (Condon 1992). Hence, the energy
imparted by spernovae cannot be radiated either directly or by transfer to
thermal particles, and is not simply building up within the disk. So it must
drive expansion of the cosmic-ray plasma: the Milky Way must be blowing a
wind that is working on the local IGM at a rate $\sim3\times10^{34}\W$.

What volume within the Local Group would this wind have filled to the
present epoch? There are two ways of answering this question. If the wind
were expanding into a vacuum, it would expand at $\sim200\kms$ and extend to
$\sim3\Mpc$ in a Hubble time. This gives us an upper limit on the radius of
influence of an $L_*$ galaxy. A lower limit comes from assuming that the
extragalactic baryons that are required by primordial nucleosynthesis theory
were once distributed in galactic halos like dark matter, and the SN-driven
wind raised its temperature by of order the virial temperature. We assume
that $\Omega_b\sim0.02$ and $\Omega_{\rm DM}\sim0.3$, so with baryons
following dark matter in a singular-isothermal halo of circular speed $v_c$,
the baryonic mass interior to radius $r$ is given by
$M_b(r)=(\Omega_b/\Omega_{\rm DM})(v_c^2/ G)r$.
Equating the energy required to heat this material by the virial
temperature to the time-integral of the SN-power, we find\footnote{By the
Tully--Fisher relation, $E_{\rm SN}\sim v_c^4$, so $r_{\rm min}$ is
independent of $v_c$.}
\begin{equation}
r_{\rm min}\sim{\Omega_{\rm DM}\over\Omega_b}
{GE_{\rm SN}\over v_c^4}\simeq180\kpc.
\end{equation}
 We see that even the present supernova rate within the Milky Way would
impact the IGM out to near the mid point to M31, and in reality the SN rate
has almost certainly been substantially larger in the past.

Recently, Tripp, Savage \& Jenkins (2000) have suggested that O$^{5+}$
absorption in quasar spectra points to a major reservoir of baryons in warm
($10^5-10^6\,$K) gas in galaxy groups. As Pen (1999) points out, the thermal
energy of this gas cannot derive from gravity alone: without
non-gravitational heating it would be so clumpy that its soft X-ray emission
would violate the constraint imposed by the unresolved component of the
X-ray  background. It is tempting to conclude that galactic winds provide
the required heat source (see below).

\section{Supernovae in galaxy clusters}

The most compelling evidence for the impact of supernovae comes from rich
clusters of galaxies. Arnaud et al (1992) show that for these objects gas
mass and $V$-band luminosity are highly correlated -- $M_{\rm gas}\sim
L_V^{1.9\pm0.3}$. At least in part the steepness of this correlation will
arise for two reasons. First, more luminous clusters tend to have higher
frations of early-type galaxies, which have lower $L_V$ per unit stellar
mass, $M_*$. Second, lower-luminosity clusters are less likely to attract
infall, and more likely to suffer outflow of gas -- Renzini (1997) finds
that systems with temperature in excess of $\sim3\kev$ do not lose gas.
Arnaud et al.\ note that $M_{\rm gas}$ is as tightly correlated with the
luminosity from E and S0 galaxies as with total luminosity: $M_{\rm gas}\sim
L_{V,{\rm E+S0}}^{1.5\pm0.25}$ and argue that the exponent in this relation
can be taken to be unity, so that with $M/L_V$ for early type galaxies set
to $\Upsilon_{{\rm E+S0}}=7\msun/L_\odot$,
 \begin{equation}\label{Mgas}
\hbox{constant}\sim {M_{\rm gas}\over M_*}\sim5.1\pm0.7\ .
\end{equation}

From the fact that $M_{\rm gas}/M_*>1$ it is clear that much of the IGM has
never been in a galaxy.

While the bulk of the gas may be primordial, the heavy elements in it are
most certainly not. With $\Upsilon_{{\rm E+S0}}=7\msun/L_\odot$ the estimate
of Arnaud et al. (1992) becomes
 \begin{equation}\label{MFeA}
{M_{\rm Fe}\over M_*}\sim(2.9\pm0.6)\times10^{-3},
\end{equation}
 while Renzini (1997) finds that for clusters hotter than $\sim2\kev$
$M_{\rm Fe}/L_B\simeq(0.02\pm0.01)\msun/L_\odot$ independent of cluster
temperature. Again adopting $\Upsilon_{{\rm E+S0}}=7\msun/L_\odot$,
Renzini's value of the Fe abundance becomes
 \begin{equation}\label{MFeR}
{M_{\rm Fe}\over M_*}\sim(2.5\pm1.5)\times10^{-3}.
\end{equation}
 The IGM-abundances of several $\alpha$-elements,  O, Ne, Mg, and especially
Si,  have been
determined for many clusters (Mushotzky et al.\ 1996). When meteoric rather
than solar-photospheric abundances are used as a point of reference, the IGM
proves to be only mildly $\alpha$-enhanced -- by a factor $\sim1.5$ relative
to solar (Brighenti \& Mathews, 1999).

What does nucleosynthesis theory have to say about  these mass fractions?
The picture is confused by uncertainties in the yields of Fe and to a
certain extent Si, from core-collapse SNe, and the rates (current and past)
of type Ia SNe.

A key input into nucleosynthetic theory is the initial mass function (IMF).
I shall assume that for $M>\msun$ this is of Salpeter's form, since there is
now significant evidence that the IMF is universal (which is surprising) and
lies near Salpeter's form at larger masses. In particular, recent work based
on Hipparcos parallaxes (Binney, Dehnen \& Bertelli, 2000) does not support
the contention of Scalo (1986) that the local IMF is steeper than Salpeter
at $M>\msun$, while Kennicut, Tamblyn \& Congdon (1994) argue that
measurements of H$\alpha$ fluxes from external disks rule out a steep IMF.

For a Salpeter IMF extending between $100$ and $0.08\msun$ we expect 0.0068
core-collapse SNe per M$_\odot$ of star formation, and $\sim1/3$ of the
initial stellar mass would by now have been returned to the ISM. The average
core-collapse SN is expected to inject $(0.1-0.14)\msun$ of Si and
$(0.07-0.14)\msun$ of Fe into the ISM (Gibson, Loewenstein \& Mushotzky,
1997), so from core-collapse SNe we expect
$\sim7.8\times10^{-4}M_{*0}=1.1\times10^{-3}M_*$ of Si and
$\sim7.1\times10^{-4}M_{*0}=1.0\times10^{-3}M_*$ of Fe, where $M_{*0}\simeq
M_*/0.7$ is the mass in stars before the onset of mass loss.

These masses fall short of those observed [eqs~(\ref{MFeA} and (\ref{MFeR})]
by a factor of order 2. The calculation ignores the existence of
intergalactic stars (Mendez et al., 1996), but it also ignores the retention
of heavy elements within galaxies. In practice these two omissions will
roughly cancel, and a more elaborate calculation would leave us
significantly short of heavy elements.

The resolution of this shortfall is controversial. Some argue for a flatter
than Salpeter IMF (David, 1997; Gibson et al., 1997). Others argue
that the missing heavies were produced by type Ia supernovae (Renzini et
al., 1993; Ishimaru \& Arimoto, 1997) -- after all more than half of the Fe
in the solar neighbourhood is thought to come from type Ia SNe.  A type
Ia SN injects $0.16\msun$ of Si and $0.74\msun$ of Fe into the ISM
(Thielemann, Nomoto \& Hashimoto, 1993), so these objects inject four times
as much Fe as Si, whereas core-collapse SNe inject at least as much Si
as Fe. Consequently, the observed ratio of Si to Fe masses in the IGM
($\sim0.8$) constrains the importance of type Ia SNe. The big
problem with imposing this constraint at the present time is uncertainty in
the ratio of mean yields of Si and Fe from core-collapse suprnovae: if
as, is perfectly plausible, core-collapse supernovae produce significantly
less Fe than Si,  one will need a good number of type Ia SNe to bring the
overall ratio $M(\hbox{Si})/M(\hbox{Fe})$ down to $\sim0.8$ as observed.

The current rate of type Ia SNe can in principle be determined in two ways:
(i) direct observation (with still depressingly uncertain results), and (ii)
from models of isolated elliptical galaxies, which require the heating rate
to decline towards the current epoch in such a way that galaxies of
appropriate luminosity and ambient IGM pressure are now suffering cooling
catastrophes in increasing numbers (Loewenstein \& Mathews, 1987; Ciotti et
al., 1991; Binney \& Tabor, 1995). The rates in times past, which will
surely differ from the present one, are well-nigh impossible to determine or
predict.

\section{SNe and the entropy of the IGM}

If rich clusters were fashioned by gravity alone, they should evolve
self-similarly.  Kaiser (1991) pointed out that it is then almost
inevitable that there should be more X-ray luminous clusters in the past
than now. This conclusion follows because the luminosity of an individual
cluster scales as the product $M\rho$ of mass times density, and with
$n$ the usual spectral index of the primoridial fluctuations, $M\sim
(1+z)^{-6/(n+3)}=(1+z)^{-3}$ for $n=-1$, while $\rho\sim(1+z)^3$, so the
characteristic luminosity will be roughly independent of $z$. The number
density of freshly collapsed objects will, on the other hand, be much higher
at high $z$. Hence there should have been many more luminous clusters in the
past, in contradiction with observation.

The obvious way of suppressing the X-ray luminosities of early clusters is
to have them form from preheated gas, which will either not be trapped by
early potential wells, or will be trapped at lower density than in the
unheated case. Kaiser estimates the entropy boost required by observing that
the gas density at the core of a rich cluster is about $10^3$ times the
current mean density, or the mean density at $z=9$.  So we could get the gas
onto the required adiabat by heating it to current cluster temperatures
$\sim10^{7.5}\,$K at $z=9$. Alternatively, we could heat it to
$[(1+z)/10]^210^{7.5}\,$K at redshift $z\ga3$.

Pen's lower limit on the entropy of gas in groups that was discussed above
implies a similar energy budget. The current maximum density of the gas is
only an order-of-magnitude greater than the current cosmic mean density, so
two order of magnitude smaller than the maximum density of cluster gas, and
the temperature would be at $\sim10^{6.5}\,$K an order-of magnitude lower
than the temperature of cluster gas. So the gas would be on a slightly
higher adiabat. On the other hand, it could be put onto that adiabat rather
later and therefore at a smaller cost in energy. 

Do SNe provide the requisite energy? The energy
release per cosmic baryon is
 \begin{equation}\label{ESN}
{E_{\rm SN}M_{\rm Fe}\over n_BM_{\rm gas}y_{\rm Fe}},
\end{equation}
 where $n_B$ is the number of baryons per unit mass, $E_{\rm SN}$ is the
energy and $y_{\rm Fe}$ is the mass of iron produced by a single SN.  Since
core-collapse SNe produce the least Fe per unit energy, we will maximize the
energy release per Fe nucleus synthesized if we assume that type Ia SNe are
unimportant. Using equations (\ref{Mgas}) and (\ref{MFeR}) to evaluate
equation (\ref{ESN}) under this assumption, and then equating the result to
the thermal energy per baryon, $\fracj32kT/\mu$, of a plasma of molecular
weight $\mu\sim0.6$, we find the plasma temperature to be
 \begin{equation}
T={2\mu\over3k}{E_{\rm SN}\over n_By_{\rm Fe}}{M_{\rm Fe}\over M_{\rm
gas}}\sim1.1\times10^7\,\hbox{K}. 
\end{equation}
 This is clearly an upper limit on the achievable temperature because it is
assumes core-collapse SNe to be dominant while ignoring radiative losses,
which are liable to be significant for many core-collapse SNe, because they
tend to go off in dense gas clouds. My judgment is that the enegy budget is
too tight, even if we assume that the gas is heated as late as $z=2$, but
others may disagree. AGN are strong candidates for providing the energy if
SNe cannot do the job (Begelman, this volume).

\section{Discussion}

As one moves down the mass scale through virial temperatures below $2\kev$,
the ratio of gas mass to stellar mass declines (Renzini, 1997). 

One interpretation of this result is  (Binney, 1980; Whitmore, Gilmore \&
Jones, 1993) that
intracluster  gas is the material from which disks are formed in the
field, and that its early enrichment to $\sim0.3Z_\odot$ by early-type
galaxies is the counterpart of the pre-enrichment of the Galactic disk by
the bulge, which Ostriker \& Thuan (1975) argued is the correct solution of
the G-dwarf problem.  Whereas in rich clusters late-infalling, high
angular-momentum gas was shocked to the virial temperature before it could
form an accretion disk around a spheroid, in systems like the Local Group
disks did form, and the floor metallicity, $\sim0.3Z_\odot$, mirrors the
current metallicity of intracluster gas.  In this picture, the declining
fraction of mass in the IGM as one moves to poorer and poorer clusters
simply reflects the fact that a larger and larger fraction of the original
IGM has settled into disks.

The argument I have given that star-forming $L_*$ galaxies blow powerful
winds suggests that all $L_*$ galaxies, cluster and field alike, lose most
of their heavy elements into deep space. If this is so, conventional galaxy
evolution models require substantial revision.  Moreover, in this picture
overwhelming quantities of metal-enriched baryons are stored in deep space.
More than one tentative line of argument now suggests that this is indeed
the case.  Within a few years we will probably know for sure.

\end{document}